\documentclass[aps,prep,preprintnumbers,amsmath,amssymb,nofootinbib]{revtex4}

\usepackage{amssymb}

\usepackage{lscape}

\usepackage{amsmath}

\usepackage{bm}

\usepackage{graphicx}

\usepackage{epsfig}

\usepackage{color}

\begin{document}

\title{Space-time waves from a collapse with a time dependent cosmological parameter}

\author{$^{1}$ Jaime Mendoza Hern\'andez\footnote{E-mail: jaime.mendoza@alumno.udg.mx}, $^{2,3}$ Mauricio Bellini
\footnote{{\bf Corresponding author}: mbellini@mdp.edu.ar}, $^{1}$ Claudia Moreno\footnote{E-mail: claudia.moreno@cucei.udg.mx} }
\address{$^1$ Departamento de F\'{\i}sica, Centro Universitario de Ciencias Exactas e Ingenier\'{\i}as, Universidad de Guadalajara
Av. Revoluci\'on 1500, Colonia Ol\'impica C.P. 44430, Guadalajara, Jalisco, M\'exico. \\
$^2$ Departamento de F\'{\i}sica, Facultad de Ciencias Exactas y
Naturales, Universidad Nacional de Mar del Plata, Funes 3350, C.P. 7600, Mar del Plata, Argentina.\\
$^3$ Instituto de Investigaciones F\'{\i}sicas de Mar del Plata (IFIMAR), \\
Consejo Nacional de Investigaciones Cient\'ificas y T\'ecnicas
(CONICET), Mar del Plata, Argentina.}

\begin{abstract}
We study the emission of space-time waves produced by back-reaction effects during a collapse of a spherically symmetric universe with a time dependent cosmological parameter, which is driven by a scalar field. As in a previous work the final state avoids the final singularity due to the fact the co-moving relativistic observer never reaches the center, because the physical time evolution
$d\tau=U_{0}\,dx^0$, decelerates for a co-moving observer which falls with the collapse. The equation of state of the system depends on the rate of the collapse, but always is positive: $0 < \omega(p) < 0.25$.
\end{abstract}

\maketitle

\section{Introduction and motivation}

The study of a collapsing system is very important when the collapse is due to the gravitational field\cite{HW}, on astrophysical scales. On the other hand, a study of a collapsing system described by a fluid with a heat flux has been treated in \cite{Sharma} and the dynamics of gravitational collapse with dissipation was studied in\cite{HS}. An interesting issue to study is the evolution of the global topology of space time during a collapse driven by a scalar field that can avoid the final singularity \cite{B,J}. Another case of great importance, but for cosmology, is the study of a cosmological collapse, where the spherical symmetry is preserved. The former system can be described by a scalar field \cite{Gundlach,GJ,G} that drives an
isotropic and homogeneous collapse. Recently, this issue was studied by the authors of this work\cite{..}, but without taking into account the cosmological parameter, which we shall consider in this work as time dependent: $\lambda(t)$. During the collapse, back-reaction effects of space-time must be considered in order to describe the system in a self-consistent manner. These fluctuations of space-time can be considered as multiple sources, statistically distributed in the universe, that produce gravitational space-time waves during the collapse. Such that effects do not must be confused with the emission of gravitational waves, which are emitted by a nonconservative quadrupolar momentum.

In this work we are interested in the study of the emission of space-time waves during the collapse of the universe in which the time is considered as scale-dependent, as well as the cosmological parameter. The manuscript is organized as follows: in Sect. II we revisit the study of boundary conditions when we variate the Einstein-Hilbert (EH) action.  In Sect. III we describe a collapse with variable timescale and a time dependent cosmological parameter $\lambda(t)$, which is driven by a scalar field $\phi$. In Sect. IV we study an example in which the collapse describes a power-law scale factor $a(t)$, and $\lambda(t)=3(\dot{a}/a)^2$. We study solutions for the waves of space-time during this collapse. Finally, in Sect. V, we develop some final comments.

\section{Physical sources from boundary conditions revisited}\label{2}

In geometrodynamics\cite{wheeler1,wheeler2} physical entities are not considered as immersed in geometry, but are regarded as manifestations of geometry, propertly. Under certain conditions, the boundary conditions must be considered in the variation of the action\cite{Y}. In the event that a manifold has a boundary
$\partial{\cal{M}}$, the action should be supplemented by a
boundary term, in order for it to be
well defined\cite{GHa}. However, there
is another way\cite{rb,rb1} to include the flux that cross a 3D-hypersurface that
encloses a physical source without the inclusion of another term
in the Einstein-Hilbert (EH) action.

\subsection{Variation of the Einstein-Hilbert action with variable cosmological constant}

We consider the variation of the Einstein-Hilbert (EH) action:
\begin{equation}
 I_{EH} =  \int d^{4}x \, \sqrt{-g} \left( \frac{R}{2\,\kappa}+\mathcal{L}_{m} \right),
\label{act}
\end{equation}
with respect to the metric tensor
\begin{equation}\label{eqn:AccionGHY}
 \delta I_{EH}  = \frac{1}{2\kappa}\,\int d^{4}x \sqrt{-g} \left[ \delta g^{\alpha \beta} \left( R_{\alpha \beta} - \frac{\bar{g}_{\alpha \beta}}{2} R + \kappa\, T_{\alpha \beta}\right) + \bar{g}^{\alpha \beta} \delta R_{\alpha \beta} \right] =0,
\end{equation}
where the stress tensor $T_{\alpha \beta}$ is defined in terms of the variation of the Lagrangian
\begin{equation}
T_{\alpha \beta} = 2\frac{\delta \mathcal{L}_{m}}{\delta g^{\alpha\beta}} - \bar{g}_{\alpha\beta}\, \mathcal{L}_{m},
\end{equation}
and describes the physical matter fields. By considering a flux $\delta \Phi$ of $\delta W^{\alpha}$ across the 3D closed hypersurface $\partial M$, we obtain that: $\bar{g}^{\alpha \beta} \delta R_{\alpha \beta} = \nabla_{\alpha}\delta W^{\alpha}=\delta \Phi$, with $\delta W^{\alpha}=\delta
\Gamma^{\alpha}_{\beta\gamma} \bar{g}^{\beta\gamma}-
\delta\Gamma^{\epsilon}_{\beta\epsilon}
\bar{g}^{\beta\alpha}=\bar{g}^{\beta\gamma} \nabla^{\alpha}
\delta\Psi_{\beta\gamma}$\cite{4}.

In order for calculate $\delta R_{\alpha \beta}$, we shall use the Palatini identity\cite{pal}
\begin{equation}
\delta{R}^{\alpha}_{\beta\gamma\alpha}=\delta{R}_{\beta\gamma}= \left(\delta\Gamma^{\alpha}_{\beta\alpha} \right)_{| \gamma} - \left(\delta\Gamma^{\alpha}_{\beta\gamma} \right)_{| \alpha},
\end{equation}
where $"|"$ denotes the covariant derivative on the extended manifold, which is described by the connections
\begin{equation}\label{ConexionWeyl}
\Gamma^{\alpha}_{\beta\gamma} = \left\{ \begin{array}{cc}  \alpha \, \\ \beta \, \gamma  \end{array} \right\} + \delta \Gamma^{\alpha}_{\beta\gamma} = \left\{ \begin{array}{cc}  \alpha \, \\ \beta \, \gamma  \end{array} \right\}+ b \,\sigma^{\alpha} g_{\beta\gamma}.
\end{equation}
The last term is a geometrical displacement $\delta \Gamma^{\alpha}_{\beta\gamma}=b \,\sigma^{\alpha}\,g_{\beta\gamma}$ with respect to the background (Riemannian) manifold, described with the Levi-Civita connections. The particular case $b=1/3$ guarantees the integrability of boundary terms in (\ref{eqn:AccionGHY}). Here, $\sigma(x^{\alpha})$ is a scalar field. In that follows we shall denote: $\sigma_{\alpha}\equiv \sigma_{,\alpha}$ as the ordinary partial derivative of $\sigma$ with respect to $x^{\alpha}$.
The condition of integrability expresses that we can assign univocally a norm to any vector in any point, so that it must be required that $\bar{g}^{\alpha\beta} \delta R_{\alpha\beta}=\nabla_{\alpha} \delta W^{\alpha}$. In the background must be fulfilled: $\Delta g_{\alpha \beta} = \bar{g}_{\alpha \beta ; \gamma} dx^{\gamma}=0$. However, on the extended manifold, we obtain
\begin{equation}\label{VariaciongWeyl}
\delta g_{\alpha \beta} = \bar{g}_{\alpha \beta | \gamma} dx^{\gamma} = - \frac{1}{3} (\sigma_{\beta} \bar{g}_{\alpha \gamma} + \sigma_{\alpha} \bar{g}_{\beta \gamma}) dx^{\gamma},
\end{equation}
where $\bar{g}_{\alpha \beta | \gamma}$ denotes the covariant derivative on the extended manifold. Therefore, the variation of the Ricci tensor on the extended manifold will be
\begin{equation}\label{VariacionRicciWeyl}
\delta R_{\alpha \beta}  = \left( \delta \Gamma^{\epsilon}_{\alpha \epsilon} \right)_{|\beta} - ( \delta \Gamma^{\epsilon}_{\alpha \beta} )_{|\epsilon} \\
 = \frac{1}{3} \left[ \nabla_{\beta} \sigma_{\alpha} + \frac{1}{3} \left( \sigma_{\alpha} \sigma_{\beta} + \sigma_{\beta} \sigma_{\alpha} \right) - \bar{g}_{\alpha \beta} \left( \nabla_{\epsilon} \sigma^{\epsilon} + \frac{2}{3} \sigma_{\nu} \sigma^{\nu} \right) \right].
\end{equation}
such that the variation of the scalar curvature is: $\delta R = \nabla_{\mu} \delta W^{\mu} = -\left[\nabla_{\mu} \sigma^{\mu} + \frac{2}{3}\sigma_{\mu} \sigma^{\mu}\right]$. Therefore, the extended Einstein tensor $\delta G_{\alpha\beta}=\delta R_{\alpha\beta} - \bar{g}_{\alpha\beta}\, \delta R$, will be
\begin{equation}\label{ein}
\delta G_{\alpha \beta}  = \frac{1}{3} \left[ \nabla_{\beta} \sigma_{\alpha} + \frac{1}{3} \left( \sigma_{\alpha} \sigma_{\beta} + \sigma_{\beta} \sigma_{\alpha} \right) + \frac{1}{2} \bar{g}_{\alpha \beta}  \nabla_{\epsilon} \sigma^{\epsilon} + \frac{1}{3} \bar{g}_{\alpha \beta} \sigma_{\nu} \sigma^{\nu} \right].
\end{equation}
Hence, if we require that $\delta G_{\alpha \beta}=-\Lambda\,\bar{g}_{\alpha\beta}$, we obtain that $\delta G=\bar{g}^{\alpha\beta}\,\delta G_{\alpha  \beta}$ is
\begin{equation}
\delta G = -4 \Lambda\left(\sigma,\sigma_{\mu}\right).
\end{equation}
Therefore, the functional $\Lambda\left(\sigma,\sigma_{\mu}\right)$ will be given by
\begin{equation}
\Lambda\left(\sigma,\sigma_{\mu}\right) = -\frac{1}{4} \left[ \Box \sigma + \frac{2}{3} \,\sigma_{\mu} \sigma^{\mu}\right].
\end{equation}
It is expected that the expectation value of $\Lambda\left(\sigma,\sigma_{\mu}\right)$ provide us the value of the cosmological parameter on the background metric.

\subsection{Extended Einstein's equations and waves of space-time}

A very important fact is that the fields $\delta \bar{W}^{\alpha}$ are invariant under gauge-transformations $\delta \bar{W}_{\alpha} = \delta W_{\alpha} - \nabla_{\alpha} \delta \Phi$, where $\delta \Phi$ satisfy $\Box \delta \Phi=0$. In this work we shall consider a collapsing universe that is isotropic and homogenous. Due to this fact, it is possible to define the redefined background Einstein's tensor
\begin{equation}\label{tr}
\bar{G}_{\alpha \beta} = G_{\alpha \beta} - \lambda(t) \,\bar{g}_{\alpha \beta},
\end{equation}
where
\begin{equation}
\lambda(t)\equiv \left<{\Lambda}\left(x^{\mu}\right)\right>=\oint\,d\Sigma\,\sqrt{-\bar{g}}\,\, {\Lambda}\left(x^{\mu}\right),
\end{equation}
such that $d\Sigma$ is the differential of the closed hypersurface given by the cyclic coordinates. They are coordinates on which the background metric tensor $\bar{g}_{\alpha\beta}$, are independent. In the case of a FRW metric with null space curvature, which is the case of interest in our present work, the cyclic coordinates are the spatial ones, so that $\left<{\Lambda}\left(x^{\mu}\right)\right>$ will be a function only of time $t$: $\lambda(t)=\left<{\Lambda}\left(x^{\mu}\right)\right>$. Because $\Lambda$ has a quantum origin, we must be more specific about the meaning of $\left<{\Lambda}\left(x^{\mu}\right)\right>\equiv \left<B\left|{\Lambda}\left(x^{\mu}\right)\right|B\right>$. It is calculated as the expectation value on a Riemann background described by the Levi-Civita symbols in (\ref{ConexionWeyl}). In our case the background quantum state can be represented in a ordinary Fock space in contrast with Loop Quantum Gravity\cite{lqg1,lqg2}, where operators are qualitatively different
from the standard quantization of gauge fields\footnote{We can define the operator
\begin{displaymath}
\delta\hat{x}^{\alpha}(t,\vec{x}) = \frac{1}{(2\pi)^{3/2}} \int d^3 k \, \check{e}^{\alpha} \left[ b_k \, \hat{x}_k(t,\vec{x}) + b^{\dagger}_k \, \hat{x}^*_k(t,\vec{x})\right],
\end{displaymath}
such that $b^{\dagger}_k$ and $b_k$ are the creation and destruction operators of space-time, such that $\left< B \left| \left[b_k,b^{\dagger}_{k'}\right]\right| B  \right> = \delta^{(3)}(\vec{k}-\vec{k'})$ and $\check{e}^{\alpha}=\epsilon^{\alpha}_{\,\,\,\,\beta\gamma\delta} \check{e}^{\beta} \check{e}^{\gamma}\check{e}^{\delta}$,
where
\begin{equation}
dx^{\alpha} \left. | B \right> =  \bar{U}^{\alpha} dS \left. | B \right>= \delta\hat{x}^{\alpha} (x^{\beta}) \left. | B \right> ,
\end{equation}
in order to
\begin{equation}
dS^2 \, \delta_{BB'}=\left( \bar{U}_{\alpha} \bar{U}^{\alpha}\right) dS^2\, \delta_{BB'} = \left< B \left|  \delta\hat{x}_{\alpha} \delta\hat{x}^{\alpha}\right| B'  \right>.
\end{equation} }.
Notice that the transformation (\ref{tr}) preserves the EH action, and
\begin{equation}\label{DinamicaMod}
\bar{G}_{\alpha \beta} = - \kappa\, \bar{T}_{\alpha \beta}.
\end{equation}
The flux that cross the 3D-gaussian hypersurface, $\delta\Phi$, is related to the cosmological parameter $\lambda(t)$, and the variation of the scalar field: $\delta\sigma$:
\begin{equation}
\delta\Phi = - \frac{4}{3} \lambda(t)\,\delta \sigma.
\end{equation}
The field ${\chi}(x^{\epsilon}) \equiv \bar{g}^{\mu\nu} {\chi}_{\mu\nu}$ is a classical scalar field, such that ${\chi}_{\mu\nu}={\delta {\Psi}_{\mu\nu}\over \delta S\,\,\,\,\,\,\,\,}$ describes the waves of space-time produced by the source through the 3D-Gaussian hypersurface \begin{equation}
\Box {\chi} = \frac{\delta \Phi}{\delta S}. \label{bb}
\end{equation}
Furthermore, the differential operator $\Box$ that acts on ${\chi}$ in (\ref{bb}), must be understanding as the wave differential operator on the background metric: $\Box\equiv \bar{g}^{\alpha\beta} \nabla_{\alpha}\nabla_{\beta}$.

\subsection{Quantization of $\sigma$}

The scalar field $\Lambda$ can be considered a functional on the extended manifold described by $\Gamma^{\alpha}_{\beta\theta}$ in (\ref{ConexionWeyl}): $\Lambda\left(\sigma,\sigma_{\alpha}\right)=-\frac{1}{4} \left( \Box \sigma+\frac{2}{3} \sigma_{\alpha} \sigma^{\alpha}  \right)$, on which behaves as a functional. By defining the action
\begin{equation}\label{AccionLambda}
\mathcal{W} = \int d^{4}x \sqrt{-\bar{g}} \hskip.1cm\Lambda(\sigma, \sigma_{\alpha}).
\end{equation}
If we require that $\delta \mathcal{W} = 0$ we obtain that $\sigma$ is a free scalar field on the extended manifold: $\Box \sigma = 0$. The scalar field $\sigma$ describes the back reaction effects which leaves invariant the action:
\begin{equation}
{\cal I}_{EH} = \int d^4 x\, \sqrt{-\bar{g}}\, \left[\frac{\bar{R}}{2\kappa} + \bar{{\cal L}}\right] = \int d^4 x\, \left[\sqrt{-\bar{g}} e^{-\frac{2}{3}\sigma}\right]\,
\left\{\left[\frac{\bar{R}}{2\kappa} + \bar{{\cal L}}\right]\,e^{\frac{2}{3}\sigma}\right\},
\end{equation}
and if we require that $\delta {\cal S}_{EH} =0$, we obtain
\begin{equation}
-\frac{\delta V}{V} = \frac{\delta \left[\frac{\bar{R}}{2\kappa} + \bar{{\cal L}}\right]}{\left[\frac{\bar{R}}{2\kappa} + \bar{{\cal L}}\right]}
= \frac{2}{3} \,\delta\sigma,
\end{equation}
where $\delta\sigma = \sigma_{\mu} dx^{\mu}$ is an exact differential and $V=\sqrt{-\bar{ g}}$ is the volume of the Riemannian manifold. The canonical geometrical momentum is $ \Pi^{\alpha}=\frac{\delta \Lambda}{\delta \sigma_{\alpha}}=-{1\over 4} \sigma^{\alpha}$, and therefore the
dynamics of $\sigma$ describes a free scalar field. Therefore, if we define the scalar invariant
$\Pi^2=\Pi_{\alpha}\Pi^{\alpha}$, we obtain that
\begin{equation}
\left[\sigma,\Pi^{2}\right] = \frac{1}{16}\left\{ \sigma_{\alpha} \left[\sigma,\sigma^{\alpha} \right]
 + \left[\sigma,\sigma_{\alpha} \right] \sigma^{\alpha} \right\}=0,
\end{equation}
hence, the relativistic quantum algebra will be given by\cite{rb,rb1}
\begin{equation}\label{con}
\left[\sigma(x),\sigma^{\alpha}(y) \right] =- i \Theta^{\alpha}\, \delta^{(4)} (x-y), \qquad \left[\sigma(x),\sigma_{\alpha}(y) \right] =
i \Theta_{\alpha}\, \delta^{(4)} (x-y),
\end{equation}
where $\Theta^{\alpha} = \hbar\, \bar{U}^{\alpha}$ and $ \Theta_{\alpha}
\Theta^{\alpha} = \hbar^2 \bar{U}_{\alpha}\, \bar{U}^{\alpha}$ for the Riemannian components of the relativistic velocity $\bar{U}^{\alpha}={dx^{\alpha}\over dS}$.

\section{Collapse with variable timescale revisited and a time dependent cosmological parameter}

As in a previous work, we shall consider a variable timescale in a collapsing spatially flat, isotropic and homogeneous universe. The line element can be represented as
\begin{equation}\label{back}
d{S}^2 = e^{-2\int \Gamma(t)\, \,dt}dt^2 - a_0^2 \,\, e^{2\int h(t) dt}\,\, {\delta}_{ij}\, \,d{x}^i d{x}^j.
\end{equation}
The metric (\ref{back}) describes the background, so that $h(t)<0$ is the collapse rate parameter on the background metric and $\Gamma(t)$ describes the time scale of the background metric. We shall use natural units: $c=\hbar=1$. The metric tensor (in cartesian coordinates), with back-reaction effects included, is
\begin{equation}\label{met1}
g_{\mu\nu} = {\rm diag}\left[\bar{g}_{00}\, e^{2\sigma/3}, \bar{g}_{11}\, e^{-2\sigma/3}, \bar{g}_{22}\, e^{-2\sigma/3},\bar{g}_{33}\, e^{-2\sigma/3}\right],
\end{equation}
which preserves the invariance of the E-H action. In order to describe a collapse, we shall consider the action for a scalar field $\phi$ which is minimally coupled to gravity
\begin{equation}\label{1}
{\cal I} = \int d^4x \, \sqrt{-\bar{g}} \,\left\{ \frac{{ \bar{R}}}{16\pi G} - \left[\frac{\dot\phi^2}{2}\,e^{2\int \Gamma(t)\,dt} - V(\phi)\right]\right\},
\end{equation}
where the collapse is driven by a scalar field, with a scalar potential $V(\phi)$. The volume of the background manifold is $\bar{v}=\sqrt{-\bar{g}}=a^3_0\,e^{-\int \Gamma(t) \,dt}\,e^{3\int H(t)\, \,dt}$. The action (\ref{1}) can be rewritten as
\begin{equation}\label{2}
{\cal I} = \int d^4x \, \sqrt{-\bar{g}}\,e^{2\int \Gamma(t)\,dt}\,\left\{ \frac{ \tilde{R}}{16\pi G} - \left[\frac{\dot\phi^2}{2} - \tilde{V}(\phi)\right]\right\},
\end{equation}
that can be considered as an action for a minimally coupled to gravity scalar field on a effective background volume $\tilde{v}=\sqrt{-\hat{g}}\,e^{2\int \Gamma(t)\,dt}$, a redefined potential $\tilde{V}(\phi)=V(\phi)\,e^{-2\int \Gamma(t)\,dt}$, and an effective scalar curvature $\tilde{R}= \bar{R}\,e^{-2\int \Gamma(t)\,dt}$. The effective volume of the background manifold in (\ref{2}), is $\tilde{v}=\sqrt{-\bar{g}} \,e^{2\int \Gamma(t)\,dt}=a^3_0\,e^{\int \Gamma(t)\, \,dt}\,e^{3\int h(t)\, \,dt}$. The dynamics of the scalar field $\phi$ is given by
\begin{equation}
\ddot\phi + \left[3 h(t)+\Gamma(t)\right] \dot\phi + \frac{\delta \bar{V}}{\delta\phi} =0. \label{infl}
\end{equation}
The background Einstein equations, are
\begin{eqnarray}
3 h^2+\lambda(t)  &=& 8\pi\, G \,\rho, \label{a} \\
-\left[3 h^2 + 2 \dot{h} + 2 \Gamma\, h + \lambda(t)\right]  &=& 8 \pi \,G\,P, \label{b}
\end{eqnarray}
where $P= \left(\frac{\dot{\phi}^2}{2}-\bar{V}(\phi)\right)\, e^{2\int \Gamma(t)\, \,dt} $ is the pressure and $\rho=\left(\frac{\dot{\phi}^2}{2} + \bar{V}(\phi)\right)\, e^{2\int \Gamma(t)\, \,dt}$ the energy density due to the scalar field. In a collapsing system, the pressure will be negative, but the kinetic component of the energy density will be significative during the evolution of the system. The equation of state that describes the dynamics of the system is:
\begin{equation}
\omega= \frac{P}{\rho} = -\left(1+\frac{2 \left(\dot{h}+ \Gamma\,h\right)}{3 h^2+\lambda}\right). \label{om}
\end{equation}
From the physical point of view, if we consider a co-moving frame where $\bar{U}_{0}=\pm \sqrt{\bar{g}_{00}}$ and $\bar{U}_j=0$, such that $j$ can take the values
$j=1,2,3$, the relativistic velocity, $\bar{U}^0={dx^0\over dS}$, will describe the rate of time suffered by a relativistic observer which is falling with the collapse of the system, for an observer which is in a non-inertial frame. Notice that the velocity will always hold the expression: $\bar{g}_{\mu\nu}\, \bar{U}^{\mu}\,\bar{U}^{\nu}=1$.

\section{Waves of space-time from the collapse}

\begin{figure}[h]
\noindent
\includegraphics[width=.9\textwidth]{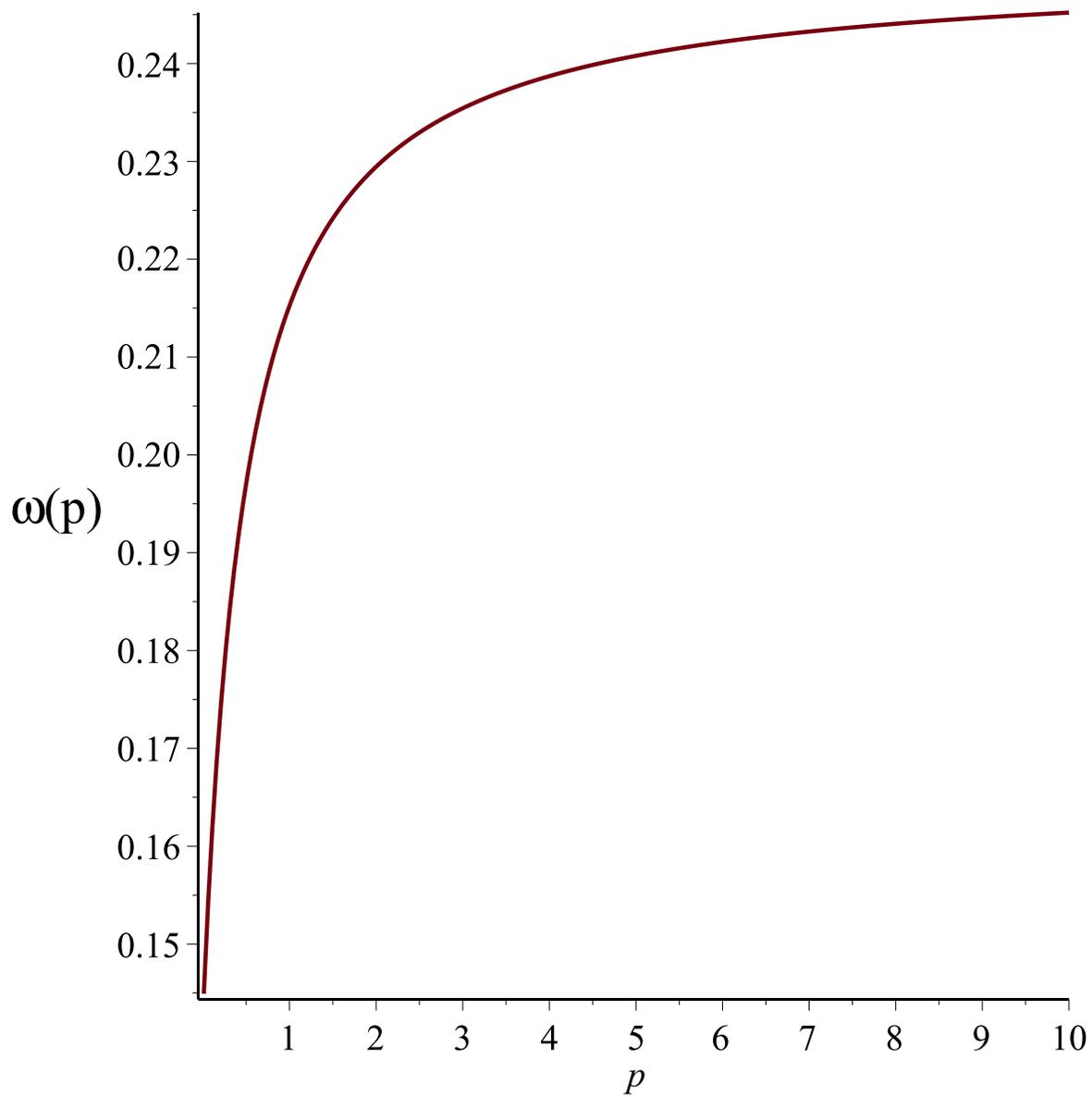}\vskip -0cm\caption{Plot of $\omega$ as a function of $p$. Notice that $\omega$ is always positive and increases with $p$, but tends to $0.25$ at $p$ tends to infinity.}\label{F1}
\end{figure}

\begin{figure}[h]
\noindent
\includegraphics[width=.9\textwidth]{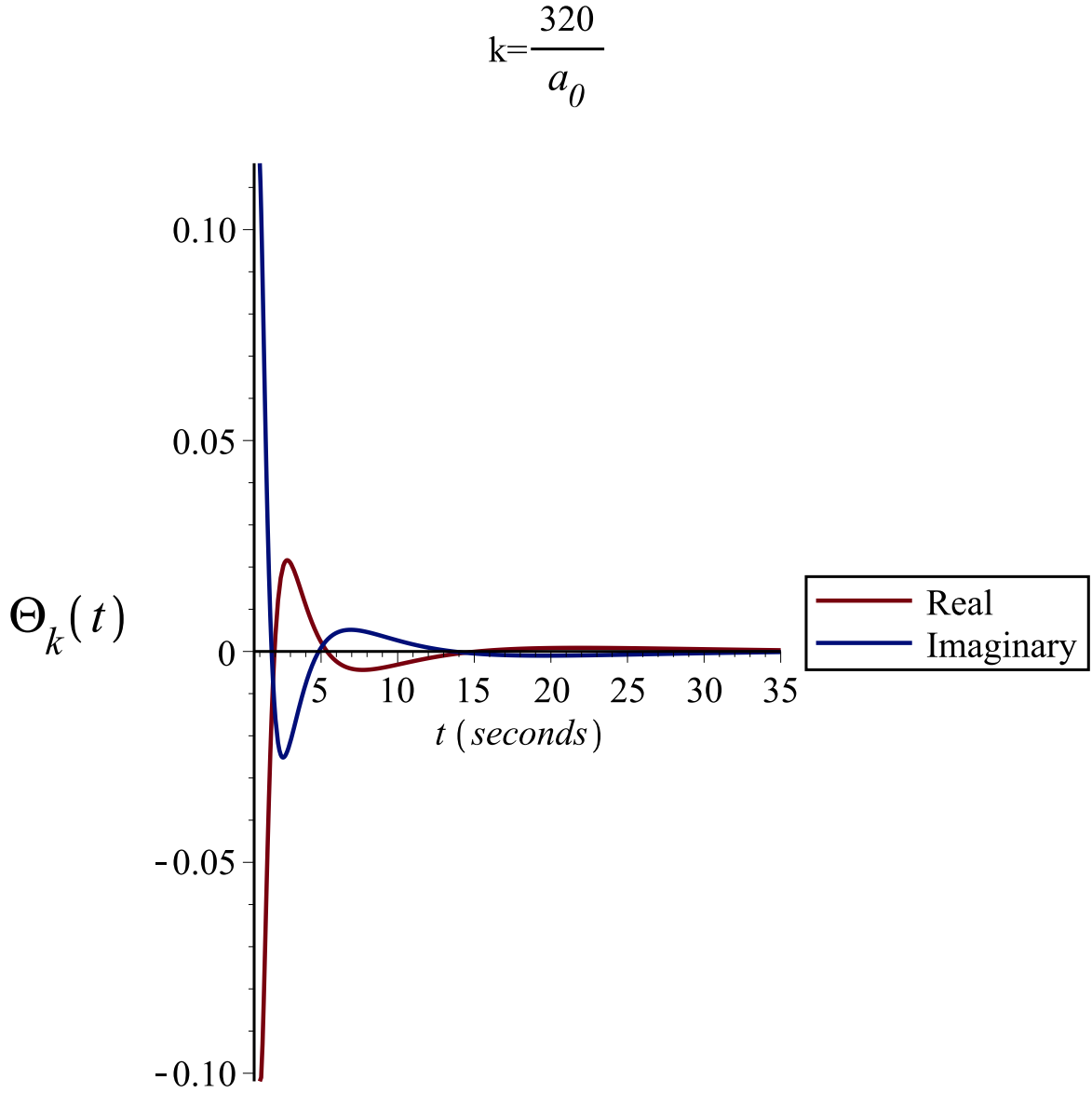}\vskip -0cm\caption{Plot of $\Theta_k(t)$ for $p=0.01$ and a wavenumber $k=320/a_0$. The red line is the real part of $\Theta_k$ and the blue one, the imaginary part. }\label{F2}
\end{figure}

\begin{figure}[h]
\noindent
\includegraphics[width=.9\textwidth]{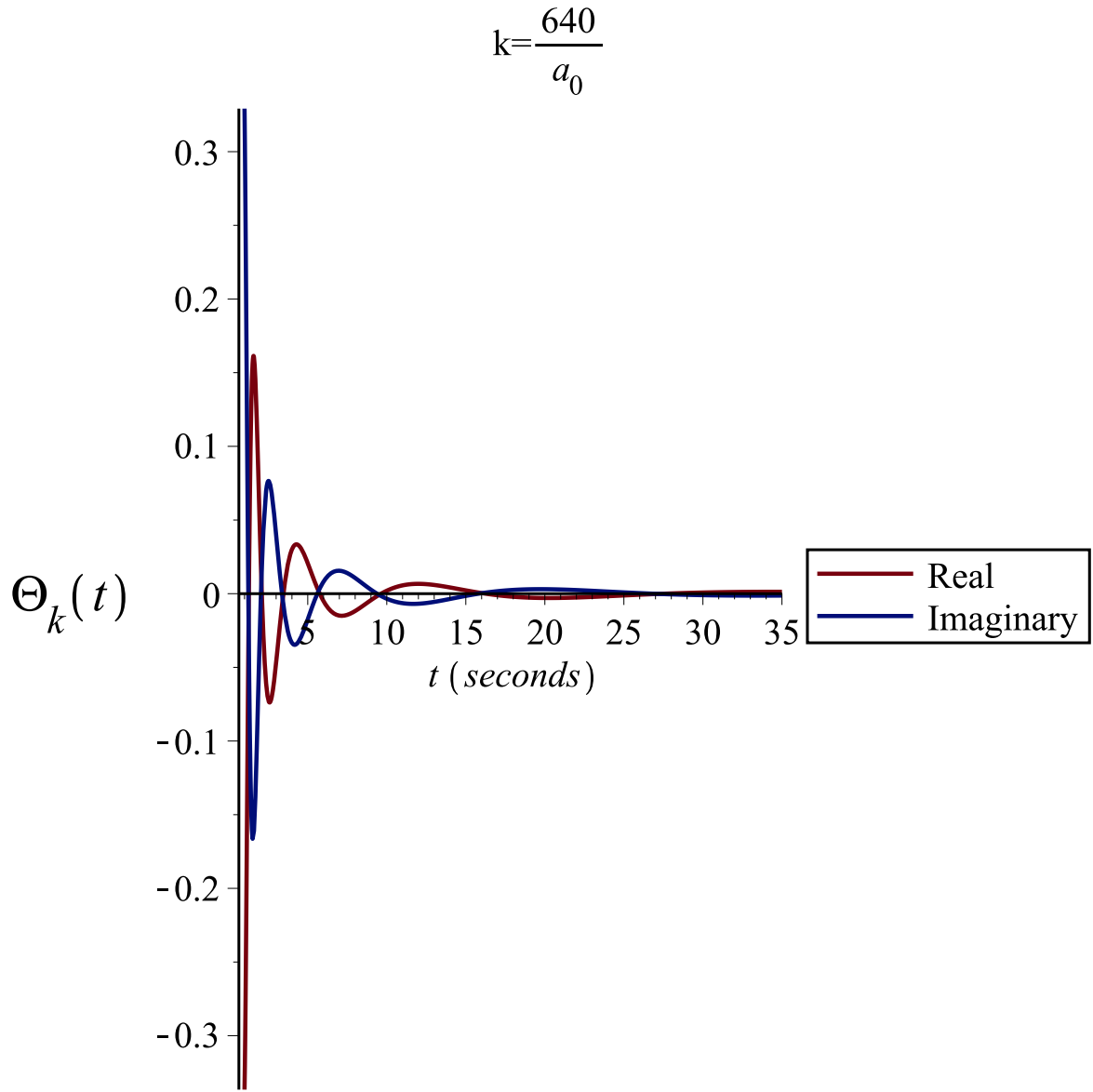}\vskip -0cm\caption{Plot of $\Theta_k(t)$ for $p=0.01$ and a wavenumber $k=640/a_0$. The red line is the real part of $\Theta_k$ and the blue one, the imaginary part. Notice that for bigger $k$ oscillates more rapidly.}\label{F3}
\end{figure}

\begin{figure}[h]
\noindent
\includegraphics[width=.9\textwidth]{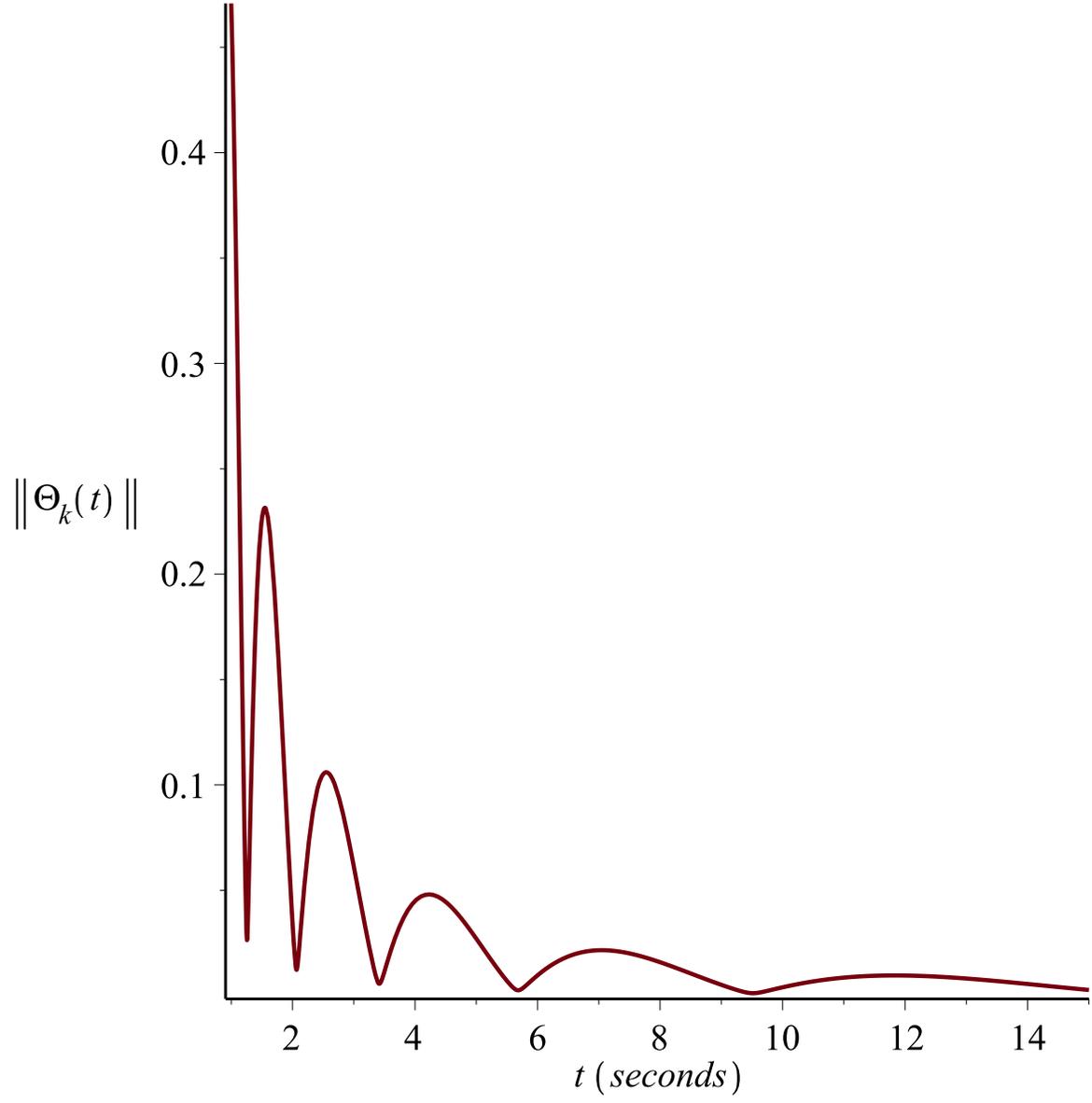}\vskip -0cm\caption{Plot of $|\Theta_k(t)|$ for $p=0.01$ and a wavenumber $k=640/a_0$. For very large times $|\Theta_k|$, stops leaping, because the frequency tends to zero due to the existence of a variable time-scale.}\label{F4}
\end{figure}

We consider the equation for the waves of space-time: $\chi$
\begin{equation}
\square {\chi}=- \frac{4}{3}\,\bar{U}^{0}\lambda(t){\dot{\sigma}},
\end{equation}
where $\bar{U}^0= \frac{x^0}{dS}= \sqrt{\bar{g}^{00}}$ is the unique relativistic nonzero component. In our model, we shall consider that $h(t)=\frac{-p}{t}$, $\lambda(t)=3h^{2}=3p^{2}/t^{2}$ and $\Gamma(t)=\frac{\alpha}{t}$. Using the Einstein equations (\ref{a}) and (\ref{b}), with the
dynamical equation for $\phi$ (\ref{infl}), we obtain that the parameter $\alpha$ can take the following values:
\begin{equation}\label{al}
\alpha_{\pm} = \frac{15p^2+p\pm \sqrt{225p^4+174p^3+49p^2}}{8p},
\end{equation}
where only the $\alpha_{+}$ is positive, so that $\alpha_{-}$ will be neglected in our treatment.
In this case the equation of state for the system is:
\begin{equation}
\omega= -\left[1+ \frac{2(1-\alpha)}{6p}\right].
\end{equation}
In the figure (\ref{F1}) we have plotted $\omega(p)$ as a function of the collapse's rate: $p>0$. Notice that $\omega >0$, and
for largest $p$-values tends to $\lim{\alpha(p)}_{p\rightarrow \infty} \rightarrow 0.25$. Furthermore, the potential and the scalar field $\phi(t)$ are respectively given by
\begin{equation}
\tilde{V}=\frac{e^{-2\int\Gamma(t)}}{8\pi G}\left[3h^{2}(t)+\dot{h}(t)+\Gamma(t)h(t)+\lambda(t)\right],
\end{equation}
\begin{equation}
\phi(t)=-\frac{1}{\alpha}\sqrt{\frac{p(\alpha-1)}{4\pi G}}\left[\left(\frac{t}{t_{0}}\right)^{-\alpha}-1\right],
\end{equation}
where, in order to $\phi$ be real we must impose that $\alpha > 1$, which implies that
\begin{equation}
p >0.
\end{equation}
Because both, ${\chi}$ and ${\sigma}$ can be expanded
into Fourier series:
\begin{eqnarray}
{\chi}(x^{\alpha}) &=& \frac{1}{(2\pi)^{3/2}} \int d^3k\,\left[A_{k}\,e^{i\vec{k}.\vec{x}}\,\Theta_k(t)+\,c.c.\right], \\
{\sigma}(x^{\alpha}) &=& \frac{1}{(2\pi)^{3/2}} \int d^3k\,\left[B_{k}\,e^{i\vec{k}.\vec{x}}\,\zeta_k(t)+\,c.c.\right],
\end{eqnarray}
where $\zeta_k(t)$ are the time dependent modes of the field ${\sigma}$, which once normalised, are:
\begin{equation}
\zeta_{k}(t)=i\sqrt{\frac{\pi t}{2pa_{0}^{3}}}\mathcal{H}_{0}^{(2)}\left[f(t)\right],
\end{equation}
where $\mathcal{H}_{\mu}^{(1,2)}[f(t)]=\mathcal{J}_{\mu}[f(t)]\pm i\,\mathcal{Y}_{\mu}[f(t)]$ are the first and second kind Hankel functions written in terms of the Bessel ones: $\mathcal{J}_{\mu}[f(t)]$ and $\mathcal{Y}_{\mu}[f(t)]$. The argument of these functions is $f(t)=\frac{k\left({t}/{t_{0}}\right)^{-2p}}{2a_0\,p}$. Therefore, once replacing $\zeta_k(t)$, we obtain the equation of motion for $\Theta_k(t)$
\begin{eqnarray}
a_{0}^{2}e^{2\int h(t)dt}\ddot{\Theta}_k(t)h(t)&+&\left(3a_{0}^{2}e^{2\int h(t)dt}(t)h(t)+a_{0}^{2}e^{2\int h(t)dt}\Gamma(t)\right)\dot{\Theta}_k(t)+k^{2}\Theta_k(t)\nonumber \\
&+& \frac{a_{0}^{2}B_{k}}{A_{k}}e^{-\int\Gamma(t)dt+2\int h(t)dt}\lambda(t)\dot{\zeta}_{k}(t)=0.
\end{eqnarray}
Therefore, the last equation will take the form
\begin{eqnarray}\label{ec}
a_{0}^{2}\left({t}/{t_{0}}\right)^{-2p}\ddot{\Theta}_k(t)&+& \left(\alpha-3p\right)
a_{0}^{2}\left({t}/{t_{0}}\right)^{-2p-1}\dot{\Theta}_k(t)+k^{2}\left({t}/{t_{0}}\right)^{-2\alpha}\Theta_k(t) \nonumber \\
&+&  \frac{{3 i\,a^2_{0}p^2\sqrt{\pi}}\,k\,B_{k}\left({t}/{t_{0}}\right)^{-\alpha-2p-5/2}}{A_{k}}
\left(\frac{\,\sqrt{2}}{4 \,\sqrt{p a^3_0}}\mathcal{H}_{0}^{(2)}
\left[f(t)\right]+\frac{1}{\sqrt{ 2p\,a^5_0}} \left({t}/{t_{0}}\right)^{-(2p)}\,\mathcal{H}_{1}^{(2)}\left[f(t)\right]\right)=0. \nonumber \\
\end{eqnarray}
In order for solve this equation we must approach in the limit case where $f(t) \gg 1$ \footnote{We use the asymptotic expressions for the Bessel functions, for $f(t) \gg 1$, which are
\begin{equation}
\mathcal{J}_{\mu}\left[A(t)\right]\approx\sqrt{\frac{2}{\pi A(t)}}\cos\left(A(t)-\frac{\mu\pi}{2}-\frac{\pi}{4}\right),
\end{equation}
\begin{equation}
\mathcal{Y}_{\mu}\left[A(t)\right]\approx\sqrt{\frac{2}{\pi A(t)}}\sin\left(A(t)-\frac{\mu\pi}{2}-\frac{\pi}{4}\right).
\end{equation}},
corresponding to small wavelengths.
The general solution for the differential equation (\ref{ec}), is:
\begin{equation}
\Theta_k(t)=\Theta_{k}(t)_{hom}+\Theta^{(p)}_{k}(t),
\end{equation}
where the homogeneous solution is
\begin{equation}
\Theta_{k}(t)_{hom}=h_{1}\left({t}/{t_{0}}\right)^{\frac{3p+1-\alpha}{2}}\mathcal{J}_{\mu}\left[A(t)\right]
+h_{2}\left({t}/{t_{0}}\right)^{\frac{3p+1-\alpha}{2}}\mathcal{Y}_{\mu}\left[A(t)\right].
\end{equation}
In order to obtain a solution to the physical problem in which the waves are produced by the source, the homogeneous solution must be null, so that we impose $h_{1}=h_{2}=0$. The particular solution, which is due to the flux on 3D-closed hypersurfaces, is
\begin{eqnarray}
\Theta^{(p)}_{k}(t)&=&C_{1}\left[\frac{(1-i)\gamma_{1}(k)\left({t}/{t_{0}}\right)^{C+E+1/2-4p}}{C+E+1/2-4p}+\frac{(1-i)\gamma_{2}(k)
\left({t}/{t_{0}}\right)^{C+E+1/2-2p}}{C+E+1/2-2p}+\frac{(1+i)\gamma_{3}(k)\left({t}/{t_{0}}\right)^{p+1+C+E+1/2-\alpha}}{p+1+C+E+1/2-\alpha}
\right. \nonumber \\
&+&\left.\frac{(1-i)\gamma_{4}(k)\left({t}/{t_{0}}\right)^{-\alpha-p+1+C+E+1/2}}{1+C+E+1/2-\alpha-p}+\frac{\left[(1+i)\gamma_{5}(k)+(1-i)\gamma_{6}(k)\right]
\left({t}/{t_{0}}\right)^{C+E+1/2}}{C+E+1/2}\right]\mathcal{J}_{\mu}[A(t)] \nonumber \\
&-& C_{2}\left[\frac{(1+i)\gamma_{1}(k)\left({t}/{t_{0}}\right)^{C+E+1/2-4p}}{C+E+1/2-4p}+\frac{(1+i)\gamma_{2}(k)\left({t}/{t_{0}}
\right)^{C+E+1/2-2p}}{C+E+1/2-2p}
+\frac{(1-i)\gamma_{3}(k)\left({t}/{t_{0}}\right)^{p+1+C+E+1/2-\alpha}}{p+1+C+E+1/2-\alpha}\right. \nonumber \\
&+&\left.\frac{(1+i)\gamma_{4}(k)\left({t}/{t_{0}}\right)^{1+C+E+1/2-\alpha-p}}{1+C+E+1/2-\alpha-p}
+\frac{\left[(-1+i)\gamma_{5}(k)+(1+i)\gamma_{6}(k)\right]\left({t}/{t_{0}}\right)^{C+E+1/2}}{C+E+1/2}\right]\mathcal{Y}_{\mu}[A(t)],
\end{eqnarray}
where the argument of the Bessel functions, is
\begin{equation}
A(t)=\frac{k\left({t}/{t_{0}}\right)^{p+1-\alpha}}{a_{0}(\alpha-p-1)},
\end{equation}
and
\begin{eqnarray}
\mu&=&\frac{\alpha-3p-1}{2(p-\alpha+1)}, \qquad C=\frac{\alpha-5p-5}{2}, \qquad D=\frac{\alpha-9p-5}{2}, \qquad E=-\alpha+3p+1, \nonumber \\
\gamma_{1}(k)&=& \frac{k^{2}}{8a_{0}^{2}p^{2}}, \qquad \gamma_{2}(k)=\frac{\pi k(\alpha-3p-1)}{4a_{0}p(\alpha-p-1)}, \qquad \gamma_{3}(k)=\frac{2k}{a_{0}(\alpha-p-1)}, \qquad \gamma_{4}(k)=\frac{k^{2}}{a_{0}^{2}p(\alpha-p-1)}, \nonumber \\
\gamma_{5}(k)&=&\frac{\pi(\alpha-3p-1)}{2(\alpha-p-1)}, \qquad \gamma_{6}(k)=\frac{\pi^{2}\alpha}{16(\alpha-p-1)^{2}}, \qquad
C_{1}=\frac{p\sqrt{a_{0}^{5}p}}{\alpha-p-1}, \qquad C_{2}= \frac{2kp\sqrt{a_{0}^{3}p}}{\alpha-p-1}.
\end{eqnarray}
In the figures (\ref{F2}) and (\ref{F3}) we have plotted the solutions $\Theta_k(t)$, for two different wavenumbers $k=320/a_0$ and $k=640/a_0$, as a function of time. Notice that in both cases the frequency tends to zero with the collapse. However, solutions with biggest wavenumbers oscillates more. In the figure (\ref{F4}) was plotted the absolute value $\left|\Theta_k(t)\right|$ for $k=640/a_0$. Notice that the amplitude of the waves of space-time decreases with time during the collapse.

\section{Final comments}

We have studied the production of space-time waves by back-reaction effects in a collapsing system, which is isotropic and homogenous and hence can be described by a scalar field. By considering the boundary conditions in the minimum action principle, we have included a time dependent cosmological parameter $\lambda(t)$. During the collapse, the equation of state is dependent of the rate of contraction $p$, but for positive $p$-values, $\omega(p)$ always remains below its asymptotic value: $0<\omega(p)<\lim \omega_{p \rightarrow \infty} \rightarrow 0.25 $. To describe the system, we have considered a time dependent cosmological parameter $\lambda(t)=3 (\dot{a}/a)^2$. Since the system is characterised by a variable time-scale $d\tau=U_{0}\,dx^0=\sqrt{\bar{g}_{00}}\,dx^0 $ that decelerates as the observer falls with the collapse, so that for a co-moving observer the final singularity is avoided never reaching the center of the sphere. During the process the amplitude of space-time waves emitted decreases with time, as well as the frequency of the waves.

\section*{Acknowledgements}

\noindent
\noindent This research was supported by the CONACyT-UDG Network Project No. 294625 "Agujeros Negros y Ondas Gravitatorias". M. B. acknowledges CONICET, Argentina (PIP 11220150100072CO) and UNMdP (EXA852/18) for financial support.

\newpage

\end{document}